\documentstyle[12pt]{article}

\begin{document}
\begin{center}
{\Large \bf The hierarchy of the preferred scales
in the fractal universe }
\bigskip

{\large D.L.~Khokhlov}
\smallskip

{\it Sumy State University, R.-Korsakov St. 2, \\
Sumy 244007, Ukraine\\
E-mail: khokhlov@cafe.sumy.ua}
\end{center}

\begin{abstract}
The fractal structure with a power index of 2 is considered
within the framework of the universe with the linear law of evolution.
The fractal structure arises due to the linear evolution of the
scale of mass.
Potential fluctuations connect two scales and thus
define the preferred scales.
The hierarchy of the preferred scales
is developed which includes universe, superclusters,
clusters of galaxies, galaxies, star clusters and stars.
\end{abstract}

\section{Introduction}

It has been suggested that the universe can be viewed
as a fractal~\cite{Pee},~\cite{Man}
where the density of the matter obeys the law
\begin{equation}
\rho \propto r^{-D}.
\label{eq:rho}
\end{equation}
If the fractal power index $D=2$,
all the objects in the universe are self-similar,
since the gravitational potential do not change with radius $r$
\begin{equation}
\varphi=\frac {Gm}{r}=G\rho r^2=const.
\label{eq:fra}
\end{equation}

Fractal galaxy distribution was discussed
in \cite{Sz}, \cite{Mar}, \cite{Luo}, \cite{Cole}.
This can be described in terms of the radial density run
\begin{equation}
N(<R) = \int_0^R dr \, \sum_{i} \delta(r-r_i) \; \propto \; R^D
\label{eq:monop}
\end{equation}
where $N(<R)$ is the average number of galaxies within radius
$R$ from any given galaxy.
The conventional point of
view \cite{Gu}, \cite{Mar1}, \cite{Lem}, \cite{Sc}
is that, on scales
$< 20\ h^{-1}\ {\rm Mpc}$, galaxies obeys $D\approx 1.2-2.2$.
On scales $> 20\ h^{-1}\ {\rm Mpc}$, the fractal power index
increases with scale towards the value $D=3$ on scales of about
$100\ h^{-1}\ {\rm Mpc}$.
On the contrary
Pietronero and collaborators~\cite{Pi}, \cite{Sy}, \cite{Ba}
claimed that galaxies have a fractal distribution
with constant $D\approx 2$ on all scales.

In the model of the universe with the linear law of evolution~\cite{Kh},
the density of the matter obeys the law~(\ref{eq:rho})
with the power index $D=2$.
Such a law arises due to the linear
evolution of the scale of mass with time.
Beneath, within the framework of this model,
the fractal structure of the universe will be considered.

\section{ The universe with the linear law of evolution }

Let us consider the model of the homogeneous and isotropic
universe~\cite{Kh} based on the premise that
the coordinate system of reference is not
defined by the matter but is a~priori specified.
Take the coordinate system of reference in the form
\begin{equation}
dl^2=a(t)^2d\tilde l^2, \quad t
\label{eq:met}
\end{equation}
where $d\tilde l^2$ is the Euclidean metric, and $t$ is
the absolute time.
The scale factor of the universe is a function of time.
Specify the evolution law of the scale factor as linear when
the scale factor grows with the velocity of light
\begin{equation}
a=ct.\label{eq:g1}
\end{equation}

Consider the universe as a particle
relative to the coordinate system of reference.
The total mass of the universe relative to
the coordinate system of reference
includes the mass of the matter and the energy of its gravity.
Adopt the total mass of the universe equal to zero,
then the mass of the matter is equal to the energy of its gravity
\begin{equation}
c^2={Gm\over{a}}.
\label{eq:o}
\end{equation}
Allowing for eq.~(\ref{eq:g1}),
from eq.~(\ref{eq:o}) it follows that
the mass of the matter changes with time as
\begin{equation}
m={c^2a\over{G}}={c^3t\over{G}},
\label{eq:p}
\end{equation}
and the density of the matter, as
\begin{equation}
\rho={{3c^2}\over{4\pi G a^2}}={3\over{4\pi G t^2}}.
\label{eq:q}
\end{equation}
So the universe with the linear law of evolution
has the fractal structure with the power index $D=2$.
The fractal structure arises due to the linear
evolution of the scale of mass with time and hence due to the
linear dependence of the scale of mass on the distance
$M\propto t\propto r$.

\section{ The permanent hierarchy of scales }

In view of eq.~(\ref{eq:q}), every distance defines its own density
\begin{equation}
\rho_i\sim r_i^{-2}.
\label{eq:rrho}
\end{equation}
The objects of the radii $r_i$ are arranged in the permanent
hierarchy of scales. This hierarchy arises due to the evolution
of the scale of mass.
Since
galaxies and clusters of galaxies approximately
obey the law~(\ref{eq:rrho}),
the formation of these
is not caused by the growth of the density fluctuations
by gravitational instability.
Since stars do not obey the law~(\ref{eq:rrho}),
it is naturally to think that
stars are formed due to the growth of the density fluctuations
by gravitational instability. In this case the radius $r_i$
defines the size of the region from which the star
forms by gravitational instability.

In view of eq.~(\ref{eq:rrho}),
the Jeans length for the region of the radius $r_i$ is given by
\begin{equation}
\lambda_{Ji}\propto\rho_{i}^{-1/2}\propto r_i.
\label{eq:lj}
\end{equation}
So all the regions of the radii $r_i$ are of scale invariance
from the viewpoint of
growth of density fluctuations by gravitational instability.

Potential fluctuations connect two scales,
the scale of homogeneity and
the scale of fluctuations
\begin{equation}
\frac{\delta M_i}{M_i}=\frac{\delta\varphi}{\varphi}.
\label{eq:DMM}
\end{equation}
Here $M_i$ is the scale of homogeneity, and
$\delta M_i$ is the scale of fluctuations.
The size of fluctuations $\delta r_i$ is given by
\begin{equation}
\frac{\delta r_i}{r_i}=\left(\frac{\delta M_i}{M_i}\right)^{1/2}.
\label{eq:drr}
\end{equation}
In the epoch of recombination $z=1400$,
the potential fluctuations are of order of
the cosmic microwave background (CMB) anisotropy
$\delta \varphi/\varphi\sim \delta T/T\sim 10^{-5}$~\cite{Ben}.
Hence the size of fluctuations is of order
$\delta r_i \sim r_i \times 10^{-2.5}$.

Before recombination, the Jeans length
is of order of the size of the region $\lambda_{Ji} \sim r_i$.
Hence the size of fluctuations is less than the Jeans length.
After recombination, the Jeans length decreases~\cite{Zeld}
and becomes of order $\lambda_{Ji} \sim r_i \times 10^{-4}$.
Hence the size of fluctuations becomes more than the Jeans length,
and the density fluctuations grow by gravitational instability.

\section{ The hierarchy of the preferred scales }

Consider the hierarchy of the preferred scales arranged
in the following way.
Let potential fluctuations connect two scales
\begin{equation}
\frac{M_i}{M_j}=\frac{\delta\varphi}{\varphi}.
\label{eq:MM}
\end{equation}
Here $M_i$ is the scale of homogeneity, and
$M_j$ is the scale of fluctuations.
$M_j$ being the scale of fluctuations relative to the scale $M_i$,
in turn, defines another scale of homogeneity.

Develop the hierarchy of the preferred scales
starting from the scale defined by the mass and
the radius of the universe.
Determine the modern age of the universe within the framework
of the universe with the linear law of evolution~\cite{Kh}.
Since density of the relativistic matter is defined by its
temperature as
\begin{equation}
\rho\sim T^{4},
\label{eq:rht}
\end{equation}
from eq.~(\ref{eq:q}) it follows that the temperature of the
relativistic matter changes with time as
\begin{equation}
T\sim a^{-1/2}\sim t^{-1/2}.\label{eq:tem}
\end{equation}
In view of eq.~(\ref{eq:tem}),
the modern age of the universe is given by
\begin{equation}
t_{0}=\alpha t_{Pl}\left(\frac{T_{Pl}}{T_{0}}\right)^2
\label{eq:age}
\end{equation}
where $\alpha$ is the electromagnetic coupling, the subscript
$Pl$ corresponds to the Planck period,
the subscript $0$ corresponds to the modern period.
Calculations yield the value
$t_0=1.06 \times 10^{18}\ {\rm s}$.
In view of eq.~(\ref{eq:p}), the mass of the universe is
$M_U=4.29 \times 10^{56}\ {\rm g}$.
This value corresponds to the relativistic matter. To transit to
the usual matter it is necessary to multiply the value by a
factor of 2.
In view of eq.~(\ref{eq:g1}),
the radius of the universe is $r_U=3.18 \times 10^{28}\ {\rm cm}$.

The potential fluctuation $\delta \varphi/\varphi$
can be determined from
the CMB spectrum.
The size of the potential fluctuation
$\Delta r$ represents the feature in the CMB spectrum.
In the fractal universe, the multipole in the CMB spectrum is
given by
\begin{equation}
\ell_{eff}=\left(\frac{\Delta r}{r}\right)^{-1}=
\left(\frac{\Delta M}{M}\right)^{-1/2}=
\left(\frac{\delta \varphi}{\varphi}\right)^{-1/2}.
\label{eq:ell}
\end{equation}
Anisotropy measurements on degree scales pin down the feature in the
CMB spectrum. The position of the feature is
$\ell_{eff}=263$~\cite{Ha},
$\ell_{eff}=260$~\cite{Li}.
Adopt the value $\ell_{eff}=260$. This corresponds to
the potential fluctuation $\delta \varphi/\varphi =1.48 \times 10^{-5}$.

With the use of the above determined mass and radius of the universe
and potential fluctuation,
develop the hierarchy of the preferred scales.
\begin{displaymath}
\begin{array}{ll}
M_1=8.6 \times 10^{56}\ {\rm g}\quad\quad &
r_1=3.2 \times 10^{28}\ {\rm cm}\\
M_2=1.3 \times 10^{52}\ {\rm g}\quad\quad &
r_2=1.2 \times 10^{26}\ {\rm cm}\\
M_3=1.9 \times 10^{47}\ {\rm g}\quad\quad &
r_3=4.7 \times 10^{23}\ {\rm cm}\\
M_4=2.8 \times 10^{42}\ {\rm g}\quad\quad &
r_4=1.8 \times 10^{21}\ {\rm cm}\\
M_5=4.1 \times 10^{37}\ {\rm g}\quad\quad &
r_5=7.0 \times 10^{18}\ {\rm cm}\\
M_6=6.1 \times 10^{32}\ {\rm g}\quad\quad &
r_6=2.7 \times 10^{16}\ {\rm cm}\\
\end{array}
\end{displaymath}
Here the second scale can be identified with superclusters,
the third scale can be identified with clusters of galaxies,
the fourth scale can be identified with galaxies,
the fifth scale can be identified with star clusters,
the sixth scale can be identified with stars.
The radius $r_6=2.7 \times 10^{16}\ {\rm cm}$
corresponds to the size of the region from which the star
forms by gravitational instability.


\begin{thebibliography}{99}

\bibitem{Pee}
P.J.E. Peebles, {\em The Large-Scale Structure of the Universe }
(Princeton University Press, Princeton, 1980)

\bibitem{Man}
B.B. Mandelbrot, {\em The Fractal Geometry of Nature }
(Freeman, New York, 1983)

\bibitem{Sz}
A.S. Szalay and D.N. Schramm, Nature {\bf 314} (1985) 718

\bibitem{Mar}
V.J.~Martinez, B.T.~Jones, R.~Dominguez-Tenreiro, and
R.V.~de~Weygaert, Ap. J. {\bf 357} (1990) 50

\bibitem{Luo}
X. Luo and D.N. Schramm, Science {\bf 256} (1992) 513

\bibitem{Cole}
P.H.~Coleman, L.~Pietronero, and R.H.~Sanders, Astron. Astroph.
{\bf 200} (1988) L32

\bibitem{Gu}
L.~Guzzo, A.~Iovino, G.~Chincarini, R.~Giovanelli, and
M.P.~Haynes, Ap. J. {\bf 382} (1991) L5-L9

\bibitem{Mar1}
V.J. Martinez and P. Coles, Ap. J. {\bf 437} (1994) 550

\bibitem{Lem}
G. Lemson and R.H. Sanders, MNRAS {\bf 252} (1991) 319

\bibitem{Sc}
R. Scaramella et al., Astron. Astroph. in press (1998),
astro-ph/9803022

\bibitem{Pi}
L. Pietronero, M. Montuori, and F. Sylos Labini, in
"Critical Dialogues in Cosmology" ed. N. Turok., in press (1997),
astro-ph/9611197

\bibitem{Sy}
F.~Sylos Labini, A.~Gabrielli, M.~Montuori, and L.~Pietronero,
Physica~A {\bf 226} (1996) 195

\bibitem{Ba}
Yu.~V.~Baryshev, F.~Sylos Labini, M.~Montuori, L.~Pietronero,
Vistas in Astronomy, Vol.~38 part 4 (1994)

\bibitem{Kh}
D.L. Khokhlov, astro-ph/9811331

\bibitem{Ben}
C.L. Benett et. al., ApJ {\bf 464} (1996) L1

\bibitem{Zeld}
Ya.B. Zeldovich and I.D. Novikov, {\em Structure and evolution of the
universe} (Nauka, Moscow, 1975, in Russian).


\bibitem{Ha}
S.~Hancock, G.~Rocha, A.~Lasenby, C.M.~Gutierrez,
MNRAS {\bf 294} (1998) L1, astro-ph/9708254

\bibitem{Li}
C.H.~Lineweaver, {\em in Proceedings of the Kyoto IAU Symposium 183:
"Cosmological Parameters and the Evolution of the Universe"},
Kyoto, Japan, August 1997, Kluwer, in press, astro-ph/9801029

\end{thebibliography}
\end{document}